\documentclass[letterpaper, 10 pt, conference]{ieeeconf}  % Comment this line out if you need a4paper

\IEEEoverridecommandlockouts                              % This command is only needed if 
\usepackage{cite}
\usepackage{amsmath,amssymb,amsfonts}
\usepackage{algorithmic}
\usepackage{graphicx}
\usepackage{graphics}
\usepackage{textcomp}
\usepackage{xcolor}
\usepackage{hyperref}
\usepackage{multirow}
\usepackage[skip=4pt]{caption}
\def\BibTeX{{\rm B\kern-.05em{\sc i\kern-.025em b}\kern-.08em
    T\kern-.1667em\lower.7ex\hbox{E}\kern-.125emX}}

\usepackage{mathptmx}
\usepackage{float}

\usepackage{enumerate}

\definecolor{shadecolor}{RGB}{240,240,240}

\title{\LARGE \bf NARX Identification using Derivative-Based Regularized Neural Networks}

\author{L.H. Peeters$^{1}$, G.I. Beintema$^{1}$, M. Forgione$^{2}$ and M. Schoukens$^{1}$% <-this % stops a space
\thanks{The activities of Marco Forgione have been supported by HASLER STIFTUNG 
under the project DEALING: DEep learning for dynamicAL systems and dynamical systems for 
deep learnING.}
\thanks{
$^{1}$Control Systems group, Department of Electrical Engineering, Eindhoven University of Technology,     The Netherlands, email: {\tt\small l.h.peeters@student.tue.nl, g.i.beintema@tue.nl, m.schoukens@tue.nl}}%
\thanks{$^{2}$IDSIA Dalle Molle Institute for Artificial Intelligence USI-SUPSI, Lugano-Viganello,                Switzerland, email: {\tt\small marco.forgione@supsi.ch}}%
}

\begin{document}

\maketitle
\thispagestyle{empty}
\pagestyle{empty}

\begin{abstract}
This work presents a novel regularization method for the identification of Nonlinear Autoregressive eXogenous (NARX) models. The regularization method promotes the exponential decay of the influence of past input samples on the current model output. This is done by penalizing the sensitivity of the NARX model simulated output with respect to the past inputs. This promotes the stability of the estimated models and improves the obtained model quality. The effectiveness of the approach is demonstrated through a simulation example, where a neural network NARX model is identified with this novel method. Moreover, it is shown that the proposed regularization approach improves the model accuracy in terms of simulation error performance compared to that of other regularization methods and model classes.
\end{abstract}
\renewcommand\thefigure{\arabic{figure}}    
\renewcommand{\theequation}{\arabic{equation}}
\renewcommand{\thetable}{\arabic{table}}

\section{introduction} \label{sec:introduction}
The goal of system identification is to estimate dynamic models that generalize well to unseen data starting from measured input-output samples. In the domain of system identification, one can distinguish between the linear and nonlinear model classes. Nonlinear behavior occurs in many engineering problems and cannot be ignored, thereby necessitating the class of nonlinear models \cite{nonlin_sys_id_road_map}. 

The Nonlinear Autoregressive with eXogenous input (NARX) model class is one of the most used nonlinear identification frameworks \cite{narx_reg_model_order_selection}. As a result, NARX models have become widely used in the domain of nonlinear system identification \cite{narx_use_rotor_bearing, narx_use_nonlin_systems_2, narx_use_nonlin_systems_3}. However, when the NARX model class is combined with general function approximators such as artificial neural networks (ANN), and since the NARX estimate minimizes the 1-step ahead prediction error, it is prone to underperform in simulation due to overfitting. Hence, it required the development of effective model selection algorithms \cite{narx_reg_model_order_selection}.

Model overfitting can be avoided by incorporating prior knowledge about the underlying system in the identification procedure. To this end, regularization approaches can be used. Regularization is a method to penalize in the estimation procedure models that do not follow the prior assumptions. In a classical learning setting, regularization is typically imposed directly on the model parameters. The most common techniques include $L_\mathit{1}$- and $L_\mathit{2}$-regularization \cite{Goodfellow2016}, by directly penalizing the magnitude of the parameters. For black-box models, such as ANNs, the interpretability of regularizing individual model parameters is often lost \cite{ann_no_parameter_interpretation}. Hence, these approaches quickly become infeasible for the inclusion of more sophisticated types of prior information. 

For some model classes, regularization methods that allow for the introduction of sophisticated priors are available. Among these is the Finite Impulse Response (FIR) model class for which e.g. the exponential decay of the impulse response is introduced as a prior by use of kernel-based regularization methods \cite{FIR_Reg_Marconato, FIR_Reg_Pilonetto}. For the NARX model class, to the authors' knowledge, no regularization method are available in the literature that allow for the introduction of prior information with such physical or system theoretical interpretations. 

In recent years, the development of expressive priors for nonlinear system identification has gained increasing attention. Most of the contributions are focused on the nonlinear FIR and Volterra model class \cite{Pillonetto2011,Birpoutsoukis2017,Stoddard2019,nfir_estimation_regularized_nns,Zancato2021}. The method recently introduced in \cite{nfir_estimation_regularized_nns} advocates a novel regularization method to incorporate prior system information without the need for parameter interpretation. More specifically, the proposed method introduces the exponential decay of the impulse response as a prior in the estimation of a Nonlinear Finite Impulse Response (NFIR) model by means of a derivative-based regularization approach. In short, this derivative-based regularization method penalizes the sensitivity of the modeled output with respect to delayed inputs. In this paper, the derivative-based regularization method in \cite{nfir_estimation_regularized_nns} is extended to the NARX case. This promotes the stability of the estimated models and improves the obtained model quality. 

The rest of this paper is organized as follows. First, regularized FIR identification is briefly revisited in Section~\ref{sec:FIR}. Next, Section~\ref{sec:narx} introduces the proposed NARX derivative-based regularization approach. Subsequently, in Section \ref{sec:results} a simulation example is presented to demonstrate the validity of the approach by comparing the results to that of different model classes. Lastly, in Section \ref{sec:conclusion} conclusions are drawn and recommendations for further research are presented.
\renewcommand\thefigure{\arabic{figure}}    
\renewcommand{\theequation}{\arabic{equation}}
\renewcommand{\thetable}{\arabic{table}}

\section{Finite Impulse Response Estimation} \label{sec:FIR}
The goal of this section is to introduce the derivative-based regularization approach by means of the relatively simple FIR model. First, in Section \ref{subsec:FIR_Models} the model structure and estimation procedure are introduced. Second, in Section \ref{subsec:FIR_Regularization} the extension of this estimation procedure with regularization is discussed. Last, in Section \ref{subsec:FIR_Derivative_based_regularization} the derivative-based regularization approach is introduced and it is shown how it can be used to incorporate the prior information of the exponentially decaying impulse response of a system in the model estimation. 

\subsection{Finite Impulse Response Models} \label{subsec:FIR_Models}
Consider the class of stable, discrete-time linear time-invariant systems. Let us assume that the system output data is corrupted by white additive Gaussian noise with zero mean and finite variance. The FIR model output is given by:
\begin{equation} \label{eq:FIR_Model}
    \hat{y}(t|\theta)=\sum_{k=0}^{n_b} g_{k} u(t-k),
\end{equation}
where $u(t)$ is the input at time $t$, $n_b$ is the number of delayed inputs and $g_k$ are the to-be-estimated impulse response coefficients, $\hat{y}(t|\theta)$ corresponds to the modeled output at time $t$ given the model parameters $\theta=\left[g_{0}, g_{1}, \ldots, g_{n_b}\right]^{\top}$. Since this is a regression task, the coefficients are typically estimated by minimization of the mean squared error cost: 
\begin{equation} \label{eq:FIR_MSE}
    V_{F}(\theta)=\frac{1}{N} \sum_{t=1}^{N}\left(y(t)-\hat{y}(t|\theta)\right)^{2}.
\end{equation}
The output data can be collected in a column vector $Y=\left[y(1), y(2), \ldots, y(N)\right]^{\top}$. Moreover, the shifted elements of the input data sequence can be organized in the regressor matrix $X$:
\begin{equation} \label{eq:FIR_DesignMatrix}
    X=\left(\begin{array}{cccc}
u(1) & u(0) & \cdots & u(-n_b+1) \\
u(2) & u(1) & \cdots & u(-n_b+2) \\
\vdots & \vdots & \ddots & \vdots \\
u(N) & u(N-1) & \cdots & u(N-n_b)
\end{array}\right).
\end{equation}
By use of the column vector $Y$, the regressor matrix $X$, and the vector of parameters $\theta$, the least squares estimate is given by:
\begin{equation} \label{eq:FIR_LS_Estimate}
    \hat{\theta}=\arg\underset{\theta}{\min } \; \|Y-X \theta\|^{2}=\left(X^{\mathrm{T}} X\right)^{-1} X^{\mathrm{T}} Y.
\end{equation}
Note that, since it is assumed that the output data is corrupted by white additive Gaussian noise,  $\hat{\theta}$ computed above also corresponds to the maximum likelihood estimate \cite{Ljung1999}.

\subsection{Regularization of FIR Models} \label{subsec:FIR_Regularization}
The regularized FIR estimate is obtained as:
\begin{equation} \label{eq:FIR_Cost_Reg}
    \begin{aligned}
    \hat{\theta}_{\text {Reg }} &=\arg\underset{\theta}{\min }\; \|Y-X \theta\|^{2}+\theta^{\mathrm{T}} R \theta  \\ 
    &= \left(X^{\mathrm{T}} X+R\right)^{-1} X^{\mathrm{T}} Y,
    \end{aligned}
\end{equation}
where $\theta^\top R \theta$ is the added regularization term, governed by the regularization matrix $R$. This regularization matrix can be used to include a prior on the exponential decay and the smoothness or bandwidth of the underlying system during the identification process \cite{FIR_Reg_Pilonetto,FIR_Reg_Marconato}. If the impulse response is modeled as a zero-mean Gaussian process, it is well known that kernel-based methods are a common way to include information about the system by using a prior covariance matrix $P$ of the parameters as:
\begin{align}
    R=\sigma^2 P^{-1}.
\end{align}
where the covariance matrix $P$ can be parametrized for instance as the diagonal-correlated (DC) $P_{\text{DC}}$ and the tuned-correlated (TC) $P_{\text{TC}}$ form \cite{Pillonetto2010,Chen2012, FIR_Reg_Pilonetto}.

\subsection{Derivative-based Regularization} \label{subsec:FIR_Derivative_based_regularization}

A key part of the prior knowledge introduced using regularized impulse response estimation is the exponentially decaying nature of the impulse response. An exponentially decaying impulse response implies that the impact of past inputs on the current output decays exponentially. Observe that for linear impulse responses an obvious equivalence exists between the partial derivative of the output with respect to the delayed inputs and the impulse response parameters:
\begin{equation} \label{eq:FIR_derivative_regularization}
    d_k(t) = \frac{\partial \hat{y}(t|\theta)}{\partial u(t-k)}=g_{k},
\end{equation}
Hence, the FIR kernel-based regularization approaches presented in \cite{Pillonetto2010, Chen2012, FIR_Reg_Pilonetto,FIR_Reg_Marconato} can be interpreted as a regularization that acts on the partial derivative of the output with respect to the delayed input, rather than the model parameters. 

This paper uses the partial derivatives as in eq.~\eqref{eq:FIR_derivative_regularization} to encode the exponentially decaying nature of a dynamic system in a more complex setting than linear impulse response estimation. More specifically, the NARX case, where the nonlinear function is represented by a feed-forward artificial neural network will be considered here. The proposed derivative-based approach enables the introduction of the exponentially decaying nature of the impulse response as a prior without the need for parameter interpretation of the considered model.
\renewcommand\thefigure{\arabic{figure}}    
\renewcommand{\theequation}{\arabic{equation}}
\renewcommand{\thetable}{\arabic{table}}

\section{NARX Identification} \label{sec:narx}
The Nonlinear AutoRegressive with eXogenous input (NARX) system class is the direct nonlinear extension of the linear time-invariant (LTI) Autoregressive with eXogenous input (ARX) system class \cite{narx_reg_model_order_selection}. Similarly to the ARX system class, delayed measured inputs and outputs are used to predict the current output. The NARX system class however uses a nonlinear function $f(\cdot)$ to map the delayed inputs and delayed outputs to the new output:
\begin{equation} \label{eq:narx}
    y(t) = f(u(t), \ldots, u(t\!-\!n_b), y(t\!-\!1),\ldots, y(t\!-\!n_a)) + e(t),
\end{equation}
where $u(t)$, $y(t)$ and $e(t)$ denote the input, noisy output, and noise disturbance at discrete time step $t$, respectively, while $n_b$ and $n_a$ correspond to number of previous inputs and outputs in the nonlinear function $f(\cdot)$. It is assumed throughout this work that the considered NARX systems are fading memory systems \cite{Boyd1985}. In this section, the estimation of a NARX model is detailed. First, the plain (unregularized) NARX model estimation procedure is briefly introduced. Next, the extension of this procedure with the derivative-based regularization method is presented.

\subsection{ANN NARX Identification} \label{subsec:NARX_ANN_Estimation}
The one-step-ahead prediction using a NARX model is given by:
\begin{equation} \label{eq:narx_pred}
    \hat{y}(t|\theta) = f_\theta(u(t), \ldots, u(t-n_b), y(t-1),\ldots, y(t-n_a)),
\end{equation}
where $\hat{y}(t|\theta)$ denotes the modeled output at time $t$, given the parameter vector $\theta$. The mean squared error of the NARX model is then given by:
 \begin{equation} \label{eq:NARX_MSE}
    \hat{\theta} = \arg\underset{\theta}{\min } \frac{1}{N} \sum_{t=1}^{N}\left[y(t) - \hat{y}(t|\theta)\right]^{2}.
\end{equation}
%where $\hat{y}(t|\theta)$ denotes NARX model output at time $t$ of the NARX model given the vector of parameters $\theta$ and $y(t)$ denotes the measured output of the real system at time $t$. 

In this work, the nonlinear function $f_\theta(\cdot)$ is parameterized as a fully connected feedforward neural network. For simplicity, a single hidden layer ANN is used throughout this paper. This neural architecture consists in the sequential connection of an input layer, a single hidden layer, and an output layer. In the NARX case, the input layer contains delayed versions of the input signal $u(t)$ and the measured output $y(t)$. The $i-\text{th}$ output $x_i(t)$ of the hidden layer is then given by: 
\begin{align}
    x_i(t)=\kappa\left(\sum_{j=1}^{n_b+1} w_{ij}^{(1,u)}u(t-j+1) + \sum_{j=1}^{n_a} w_{ij}^{(1,y)}y(t-j) + w_{i0}^{(1)}\right),
    \label{eq:out_mlp}
\end{align}
where $w_{i0}^{(1)}$ is the bias of neuron $i$, and $w_{ij}^{(1,u)}$, $w_{ij}^{(1,y)}$ are the weights related to the past inputs and outputs respectively, and $\kappa(\cdot)$ is the activation function. Next, the outputs of the hidden layer~\eqref{eq:out_mlp} are transformed by a linear output layer into the NARX model output given by:
\begin{align}
    \hat{y}(t|\theta)=\sum_{i=1}^Q w_{i}^{(2)}x_i(t) + w_{0}^{(2)},
    \label{eq:out_nn}
\end{align}
where $w_{i}^{(2)}$ is the $i-\text{th}$ weight of the output neuron, $w_{0}^{(2)}$ is the bias of the output, and $Q$ is the number of neurons in the hidden layer. The rectified linear unit, hyperbolic tangent function, sigmoid function and tansig function are amongst the most commonly used activation functions used for $\kappa(\cdot)$. The network weights and biases are grouped in the parameter vector $\theta$.

Due to the nonlinear parameterization of the function $f_\theta(\cdot)$ in Equation \ref{eq:narx_pred}, no analytical solution exists to the optimization problem in Equation \ref{eq:NARX_MSE}. Thus, iterative 
numerical optimization methods such as gradient descent approaches are used to minimize the cost function. 

The lack of generalization to unseen data, or overfitting, is a common problem in data-driven modelling. Overfitted models tend to memorize all the data, including unavoidable noise in the training set, instead of learning the system dynamics hidden behind the data \cite{NN_Overfitting}. To reduce the effect of overfitting, regularization techniques such as early stopping (ES), network-reduction and weight decay ($L_\mathit{2}$-regularization) are currently used.

\subsection{Derivative-based Regularization for NARX Identification} \label{subsec:NARX_ANN_Regularization}
This section introduces the novel derivative-based regularization approach for the identification of NARX models. Since the optimization problem~\eqref{eq:NARX_MSE} is nonlinear in the parameters, including an exponentially decaying prior in the identification process is not straightforward. This section proposes to achieve this by making use of the partial derivatives of the output with respect to the delayed inputs:
\begin{equation} \label{eq:NARX_Cost_Simulation}
\begin{aligned}
V_{N_{Reg}}(\theta)=\frac{1}{N} \sum_{t=1}^{N}\left[y(t) - \hat{y}(t|t)\right]^{2} + \gamma \sum_{k=0}^{T} \psi_k^{\top} R_{k} \psi_k, \\
\psi_k=\left[\begin{array}{c}
    d_{k}^*(1) \\
    d_{k}^*(2) \\
    \vdots \\
    d_{k}^*(N)
    \end{array}\right], \quad d_k^*(t) = \frac{\partial \hat{y}(t+T|t)}{\partial u(t+T-k)}, \quad R_k = \alpha^{-k} I_N,
\end{aligned}
\end{equation}
where $d_{k}^*(t)$ corresponds to the partial derivative of the $T$-step ahead simulation with respect to the $k$-step delayed input at time $t$. The dependency of $\hat{y}(t|t)$ on $\theta$ is dropped for compactness of notation.

The $T$-step ahead simulation of the NARX model is used to compute the required partial derivatives. 
%This prediction is obtained by re-using the predicted model outputs in the NARX model while neglecting the effects of the disturbing noise, resulting in:
%\begin{equation} \label{eq:NARX_T_step_ahead}
%    \hat{y}(t+T|t) = F(u(t+T), \ldots, u(t-n_b), y(t-1),\ldots, y(t-n_a)),
%\end{equation}
Specifically, the simulation is performed by injecting the \emph{predicted} model outputs in the NARX model (thus neglecting the effects of the disturbing noise) resulting in:% This results in:
\begin{equation} \label{eq:NARX_T_step_ahead}
    \hat{y}(t+T|t-1) = F(u(t+T), \ldots, u(t-n_b), y(t-1),\ldots, y(t-n_a)),
\end{equation}
where the nonlinear function $F(\cdot)$ is a composition of the recursively used nonlinear function $f(\cdot)$ of the NARX model presented in \eqref{eq:narx_pred}. Note that this does not result in the exact T-step ahead prediction output as higher order noise effects are ignored. This has, however, no significant impact unless extremely low signal-to-noise regimes are considered \cite{Khandelwal2018}. 
The simulation window $T$ should be chosen larger than the slowest time constant of the system such that the measured outputs $y(t-1),\ldots, y(t-n_a)$ have little impact on the $T$-step ahead simulation $\hat{y}(t+T|t)$, but shouldn't be chosen too large as it slows down the parameter optimization. Note that $F$ simplifies to the classical 1-step ahead prediction for $T=1$.

Thus, to calculate the regularization term, first for each time step $t$ the NARX model is used in closed-loop simulation to obtain the $T$-step-ahead simulation $\hat{y}(t+T|t)$. To do this, the model predictions are recursively injected. Then, for this $T$-step-ahead prediction the partial derivative with respect to the delayed inputs is calculated. These partial derivatives are then penalized exponentially for a growing delay (by the factor $\alpha^{-k}$, $\alpha \in [0,1]$) to incorporate the prior of the exponential decay of the impulse response.

\subsection{Cost Function Optimization}
The optimization problem outlined in \eqref{eq:NARX_Cost_Simulation} is minimized using the Adam optimizer stochastic gradient descent algorithm. The gradients as well as the  partial derivatives $d_k^\star(t)$ are obtained using reverse-mode automatic differentiation.
%the backpropagation automatic differentiation algorithm.

\renewcommand\thefigure{\arabic{figure}}    
\renewcommand{\theequation}{\arabic{equation}}
\renewcommand{\thetable}{\arabic{table}}

\section{Results} \label{sec:results}
This section demonstrates the effectiveness of the described regularization method. To this end a performance comparison is presented for the NARX model. More specifically, the NARX model estimation with and without the described regularization method are considered. In addition, it is also compared to the results obtained using a nonlinear output-error (NOE) model and an LTI FIR model. The performance comparison is made by use of a simulation example as presented in Section \ref{subsec:Simulation_Example}. The results are presented in Section \ref{subsec:comparison}. At last, the effectiveness of the derivative-based regularization approach is discussed in Section \ref{subsec:effectiveness}.

\subsection{Data Generation} \label{subsec:Simulation_Example} 
\begin{figure}[H]
    \centering
    \includegraphics[width=\columnwidth]{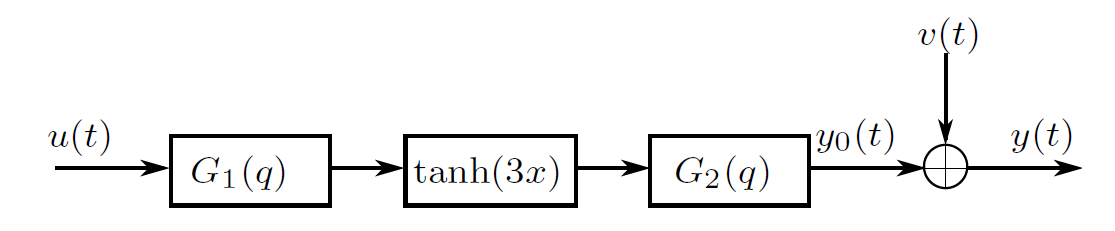}
    \caption{Wiener-Hammerstein system for data generation \cite{nfir_estimation_regularized_nns}.}
    \label{fig:WH_System}
\end{figure}

To demonstrate the effectiveness of the approach a simulation example is used. The considered system is a Wiener-Hammerstein system as presented in Figure \ref{fig:WH_System}. 
In Figure \ref{fig:WH_System}, $u(t)$ is the input of the system, $y_0(t)$ is the noiseless output of the system, $v(t)$ corresponds to the noise signal and $y(t)$ is the noisy output i.e. $y(t) = y_0(t) + v(t)$. The \texttt{tanh()}-block is used to introduce static non-linearity in the system and the blocks $G_1(q)$ and $G_2(q)$ model the system dynamics:
\begin{equation} \label{eq:WHSystem_TransferFunctions}
    \begin{aligned}
    G_{1}(q)&=\frac{0.0451+0.0902 q^{-1}+0.0451 q^{-2}}{1-1.3860 q^{-1}+0.7069 q^{-2}}, \\
    G_{2}(q)&=\frac{0.2545+0.0073 q^{-1}+0.0073 q^{-2}+0.2545 q^{-3}}{1-1.1495 q^{-1}+0.7459 q^{-2}-0.0729 q^{-3}},
    \end{aligned}
\end{equation}
where $q^{-1}$ corresponds to the delay operator. The input $u(t)$ is a zero-mean, unit variance multisine signal with random phase \cite{Multisine_input}. 
For the training set $N=1024$ samples are used and the input signal excites the full frequency range up to $f_s/2$ using a flat amplitude spectrum. The noise signal $v(t)$ is zero-mean Gaussian with a standard deviation $\sigma_v = 0.01$. The training set is split in a 80/20-ratio to obtain the training and validation set that are used for the model estimation.

For testing the performance, a white and colored test set are considered. Both the white and colored test set consist of $N=10000$ samples and are noiseless. For the white test set, the input signal excites the full frequency range up to $f_s/2$ using a flat amplitude spectrum. For the colored test set the input signal excites the full frequency range up to $f_s/10$.

\subsection{Comparison} \label{subsec:comparison} 
This section presents the considered performance metric, the model architectures and the obtained results. The performance metric to assess the simulation performance of the models is the normalized root-mean-squared-error (NRMSE): 
\begin{equation} \label{eq:performance_metric}
    S_{NRMSE} = \frac{\sqrt{\frac{1}{n} \sum_{t=1}^{n}\left[y(t) - \hat{y}(t|1)\right]^{2}}}{\sigma_y^2},
\end{equation}
where $n$ corresponds to the length of the sequence of output data $y$. Furthermore, $\sigma_y^2$ corresponds to the variance of the output sequence and $\hat{y}(t|1)$ is the simulated output of the considered model at time $t$.

As discussed in Section III.A, the nonlinear function of the NARX and NOE model an ANN representation is chosen since they can uniformly approximate any continuous function \cite{narx_approximate_function1, narx_approximate_function2}. The NARX ANN model, referred to as the NARX network, is a single-hidden-layer feed-forward network. In the hidden layer the tanh activation function is used. The output layer consists of a single neuron with a linear activation function. For the comparison of the models two cases with differing model orders are considered: 
\begin{enumerate}
    \item \noindent \textbf{High Model Order}: $Q = 20$ neurons in the hidden layer and $n_b = n_a = 30$ delayed inputs and outputs. The number of neurons and considered delayed inputs and outputs is chosen sufficiently large. This way, the model is able to accurately model the system. This case, referred to as the high model order (HMO) case, aims to highlight the regularization capabilities of the proposed algorithm.
    \item \noindent \textbf{Optimized Model Order}: $Q = 10$ neurons in the hidden layer and $n_b = n_a = 15$ delayed inputs and outputs. The number of neurons and considered delayed inputs and outputs is determined by a model order scan. For this model order scan, a NARX network with no regularization (only early stopping is used) was trained. The NARX network that had the best normalized-root-mean-square error (NRMSE) in simulation \eqref{eq:performance_metric} on the test dataset was chosen. This case is referred to as the optimized model order (OMO) case. Note that the test dataset is abused as a validation set during the model selection procedure. This acts as a possible advantage for the classical NARX identification approach on which this model order selection is carried out when evaluating the test performance.
\end{enumerate} 

For both model orders, four different modeling approaches are considered:

\begin{enumerate}
    \item \noindent \textbf{LTI}: a nonregularized linear time-invariant finite impulse response model estimated with the \texttt{impulseest}-module in Python 3 \cite{fiorio2021impulseest}. For this model the estimated number of impulse response coefficients corresponds to the number of delayed inputs $n_b$ as presented above.
    \item \noindent \textbf{ES}: a NARX network without the derivative-based regularization method applied. Early stopping (ES) is applied using the 1-step ahead prediction error (same criterion as for training) on the validation dataset with a patience of 1000 epochs.
    \item \noindent \textbf{DR}: a NARX network with the derivative-based regularization method applied. For this model, the regularization-hyperparameters ($\alpha, \gamma$) are estimated by means of a 10x10 grid search using the validation dataset. For $\alpha$ the linearly spaced grid ranges from 0.60 to 0.75 and for $\gamma$ the logarithmically spaced grid ranges from $5 \times 10^{-7}$ to $5 \times 10^{-3}$. The simulation depth for this model estimation is set to $T = 50$. Early stopping is applied using the 1-step ahead prediction error (same criterion as for training) with a patience of 1000 epochs.
    \item \noindent \textbf{NOE}: a nonlinear output error (NOE) model. This model is trained directly by minimization of the simulation error. Early stopping is applied using the simulation error (same criterion as for training) with a patience of 1000 epochs.
\end{enumerate}

\begin{figure}[htb]
    \centering
    \includegraphics[width= \columnwidth]{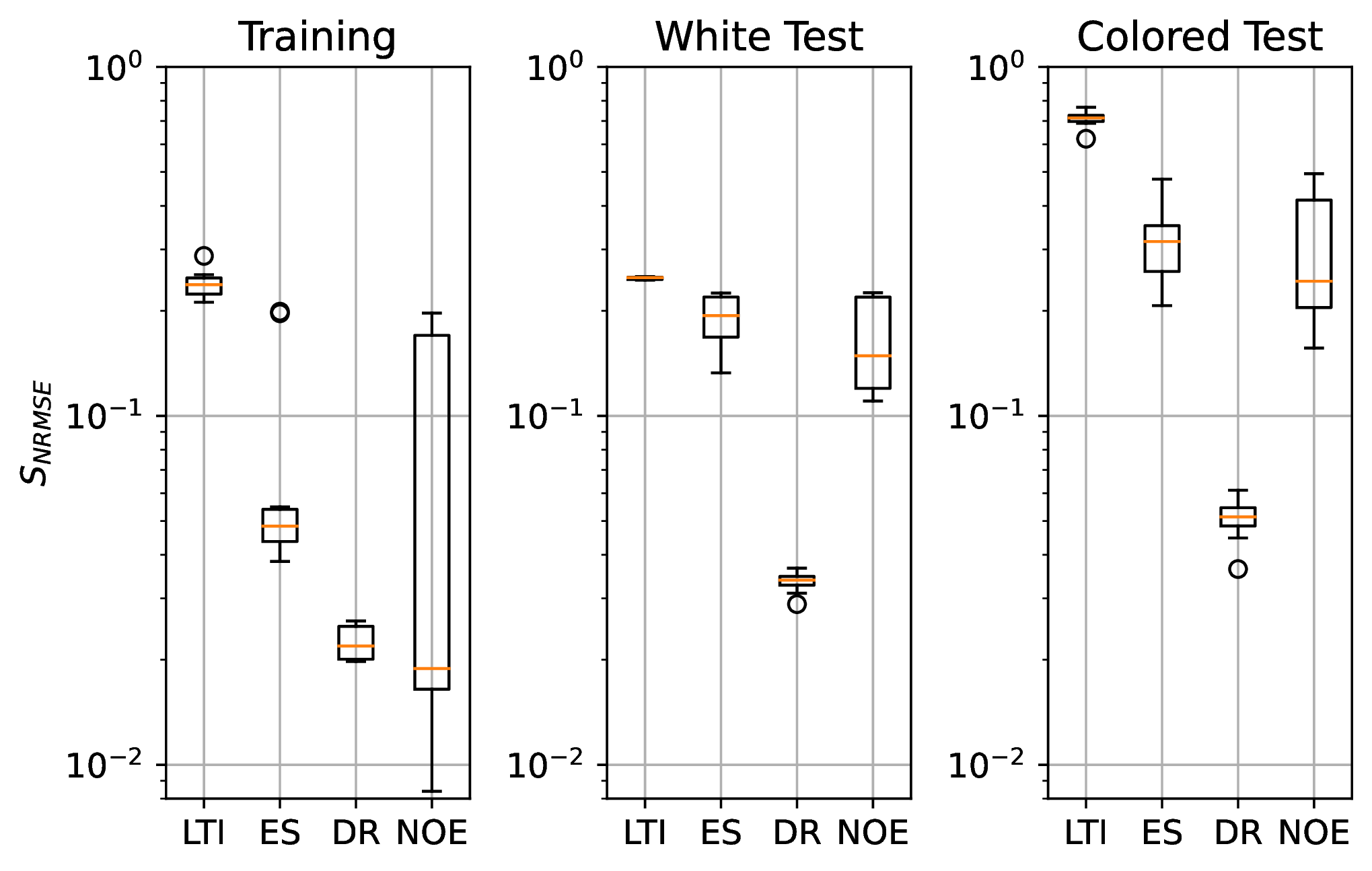}
    \caption{Boxplot of the simulation NRMSE for the 10 Monte-Carlo runs for the HMO-case on the training, white test and colored test data. }
    \label{fig:boxplot_HMO}
\end{figure}

\begin{figure}[htb]
    \centering
    \includegraphics[width= \columnwidth]{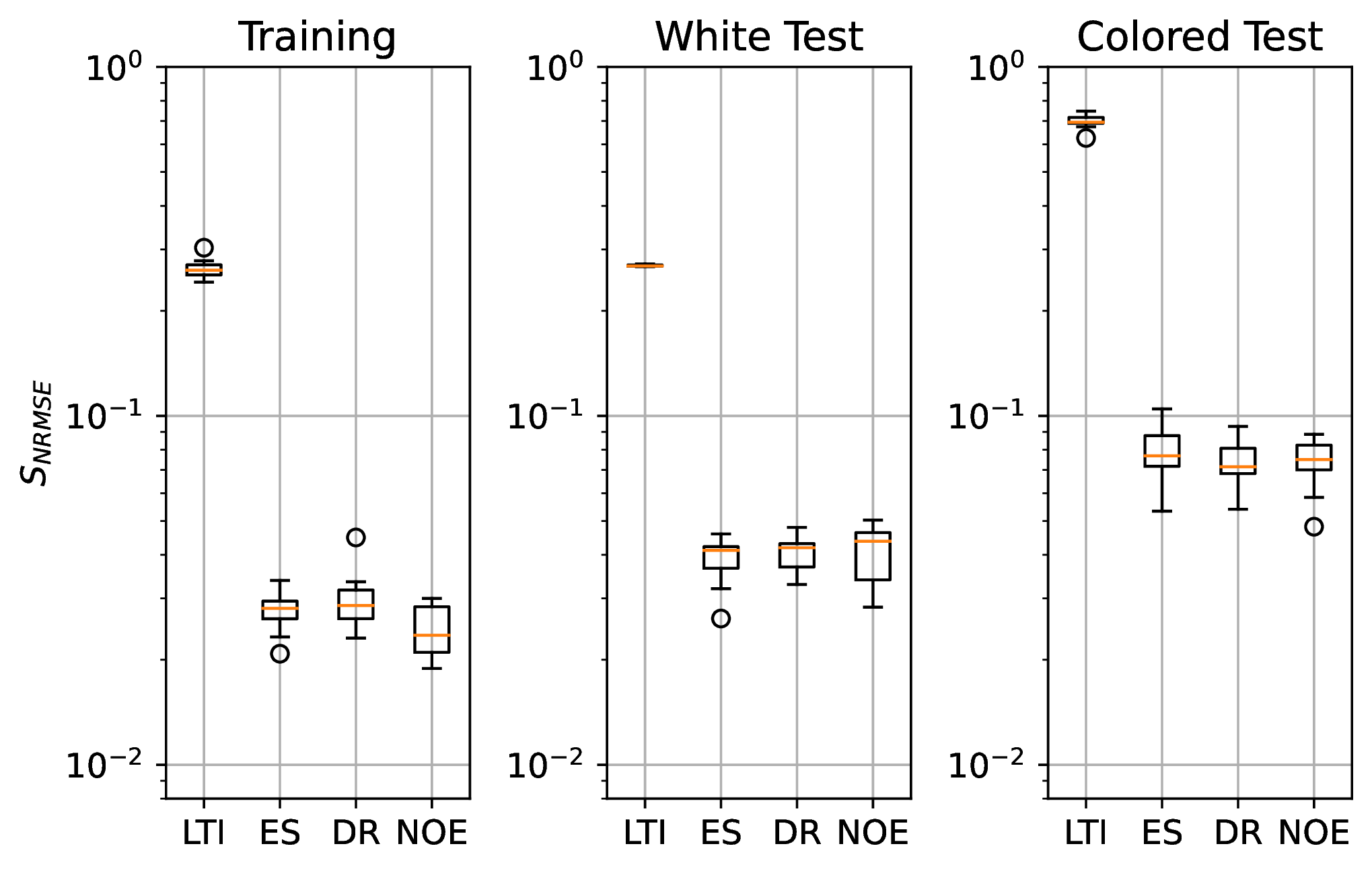}
    \caption{Boxplot of the simulation NRMSE for the 10 Monte-Carlo runs for the OMO-case on the training, white test and colored test data. }
    \label{fig:boxplot_OMO}
\end{figure}

The ES, DR and NOE models are estimated using a Pytorch implementation with a maximum of 10000 epochs and batch size equal to 1024. The default Pytorch ADAM optimizer settings are used. For the ES, DR and NOE case the model that performs best on the validation set is used for comparison of results. To provide statistically reliable results, a Monte Carlo simulation of 10 runs is performed for both model order cases. The results of the Monte-Carlo simulation for all considered models are presented in Tables \ref{tab:comparison_results_osa} and \ref{tab:comparison_results} and Figures \ref{fig:boxplot_HMO} and \ref{fig:boxplot_OMO}.

\begin{table}[htb]
\caption{Median NRMSE in one-step ahead prediction over 10 Monte-Carlo simulations for all considered models.}
\label{tab:comparison_results_osa}
\centering
\resizebox{0.9 \columnwidth}{!}{\begin{tabular}{r|c|l|l|l}
\multicolumn{1}{l|}{} & \multicolumn{1}{c|}{LTI} & \multicolumn{1}{c|}{ES} & \multicolumn{1}{c|}{DR} & \multicolumn{1}{c}{NOE} \\ \hline \hline
\multicolumn{1}{l|}{\textbf{HMO}} &  &  &  &  \\ 
Training & / & \textbf{0.016} & 0.017 & 0.021 \\ 
White Test & / & 0.137 & \textbf{0.025} & 0.109 \\ 
Colored Test & / & 0.206 & \textbf{0.038} & 0.182 \\ \hline \hline
\multicolumn{1}{l|}{\textbf{OMO}} &  &  &  &  \\ 
Training & / & 0.021 & 0.021 & 0.026 \\ 
White Test & / & 0.027 & 0.028 & 0.032 \\ 
Colored Test & / & 0.053 & 0.046 & 0.049 \\ 
\end{tabular}}
\end{table}

\begin{table}[htb]
\caption{Median NRMSE in simulation over 10 Monte-Carlo simulations for all considered models.}
\label{tab:comparison_results}
\centering
\resizebox{0.9 \columnwidth}{!}{\begin{tabular}{r|l|l|l|l}
\multicolumn{1}{l|}{} & \multicolumn{1}{c|}{LTI} & \multicolumn{1}{c|}{ES} & \multicolumn{1}{c|}{DR} & \multicolumn{1}{c}{NOE} \\ \hline \hline
\multicolumn{1}{l|}{\textbf{HMO}} &  &  &  &  \\ 
Training & 0.238 & 0.048 & 0.022 & \textbf{0.019} \\ 
White Test & 0.249 & 0.194 & \textbf{0.034} & 0.149 \\ 
Colored Test & 0.714 & 0.316 & \textbf{0.051} & 0.243 \\ \hline \hline
\multicolumn{1}{l|}{\textbf{OMO}} &  &  &  &  \\ 
Training & 0.262 & 0.028 & 0.029 & 0.024 \\ 
White Test & 0.269 & 0.041 & 0.042 & 0.044 \\ 
Colored Test & 0.694 & 0.077 & 0.071 & 0.075 \\ 
\end{tabular}}
\end{table}

The median one-step ahead prediction NRMSE and the simulation NRMSE are reported in Table~\ref{tab:comparison_results_osa} and Table~\ref{tab:comparison_results} respectively. Although the proposed approach has a higher one-step ahead prediction and simulation error than the ES and NOE models respectively, it outperforms both approaches on the test datasets. This is most apparent from the obtained performance on the colored test set, which is differs most from the training dataset. This is also confirmed by the boxplots of the 10 different Monte-Carlo runs depicted in Figure~\ref{fig:boxplot_HMO} and \ref{fig:boxplot_OMO}. These figures also visualize that the ES- and NOE-model, despite using early stopping during model estimation, do not generalize well to the simulation task on the test data. This indicates that an early-stopping regularization approach is insufficient in this regard. For the OMO-case it can be seen that the proposed method (DR) performs similarly compared to the other two nonlinear models (ES \& NOE). Finally, note that the best overall model is obtained using the proposed regularization approach on the HMO case. Hence, providing sufficient flexibility to the NARX model, and in a second step managing this complexity through the proposed regularization approach outperforms classical model order selection.

\subsection{Regularization Impact} \label{subsec:effectiveness}
\begin{figure}[H]
    \centering
    \includegraphics[width= \columnwidth]{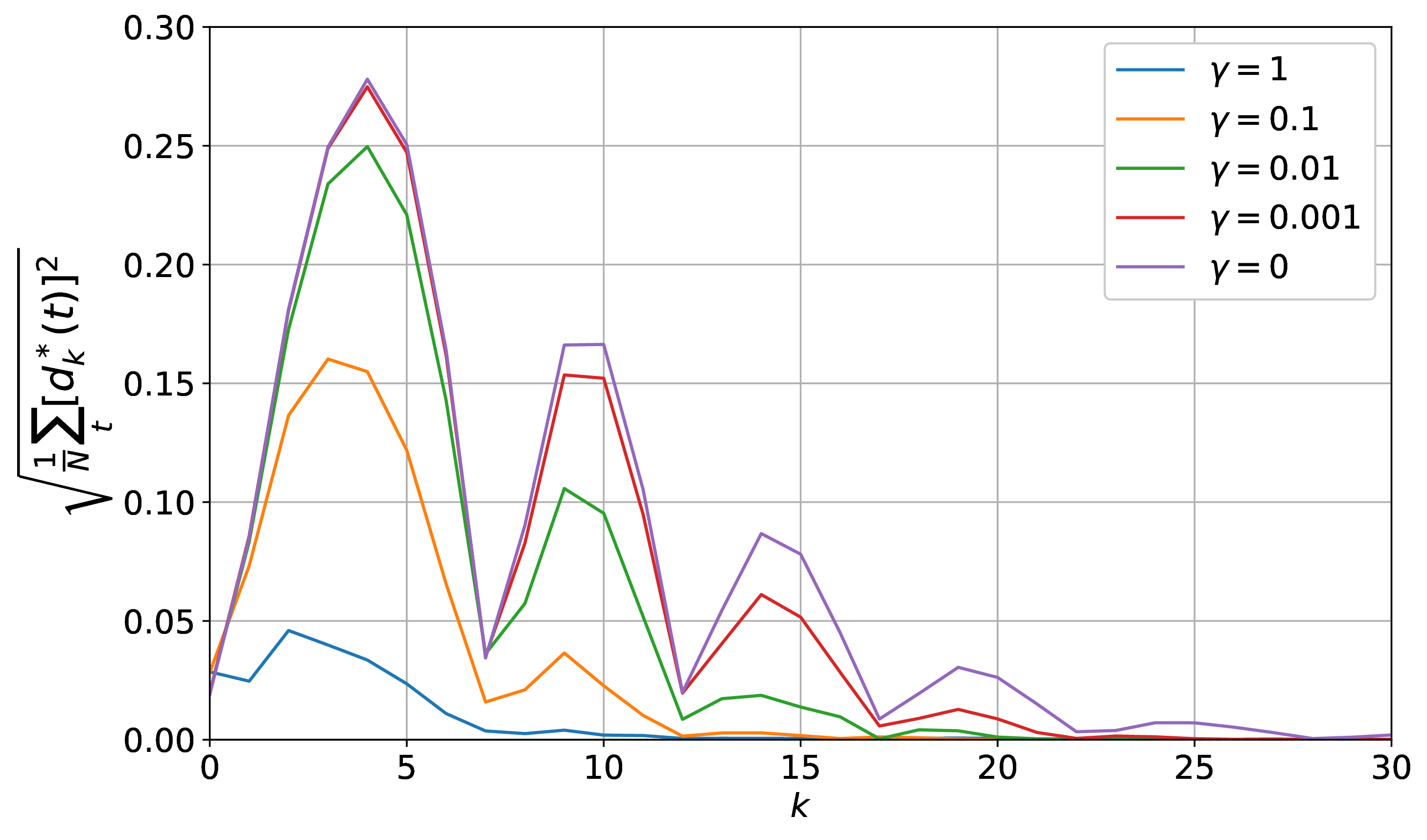}
    \caption{Partial derivative of the modeled output w.r.t. delayed inputs evaluated on the training set for the HMO-DR model case.}
    \label{fig:decay_plot}
\end{figure}

In this section, the impact of the proposed regularization approach is briefly discussed. To this end, Figure~\ref{fig:decay_plot} shows the partial derivative of the $T$-step ahead model prediction with respect to the delayed inputs. From the figure it can be seen that the effect of delayed inputs on the output decays exponentially. Moreover, as expected, the sensitivity of the delayed inputs on the modeled output decreases for growing $\gamma$ values. This indicates the proposed regularization method can be used to incorporate prior information about the exponential decay of the impulse response of the system in the NARX network model estimation. 

%In addition note that the bounces in the figure connect to the theory about linearization of the impulse response over the simulation trajectory as presented in Section \ref{subsec:NARX_ANN_Regularization}. 
\renewcommand\thefigure{\arabic{figure}}    
\renewcommand{\theequation}{\arabic{equation}}
\renewcommand{\thetable}{\arabic{table}}

\section{Conclusion} \label{sec:conclusion}
This paper introduced a novel regularization approach for NARX identification. This novel approach introduces a prior on the exponential decay of the influence of past inputs on the current model output in the estimation problem. This is an extension on the exponentially decaying prior used in regularized impulse response estimation approaches. It does this by penalizing the sensitivity (partial derivative) of the output with respect to delayed inputs. The presented results demonstrate the promising performance of the proposed approach, the generalization from the training dataset to test dataset (both with similar and different spectral properties) is improved drastically. The novel, regularized, NARX identification outperforms both classical NARX and NOE identification approaches.

\bibliographystyle{./bibliography/IEEEtran}
\bibliography{./bibliography/IEEEabrv.bib,./bibliography/IEEEexample.bib}

\renewcommand\thefigure{\arabic{figure}}    
\renewcommand{\theequation}{\arabic{equation}}
\renewcommand{\thetable}{\arabic{table}}

\appendix \label{sec:appendix}

This appendix illustrates how the function $F(.)$ in \eqref{eq:NARX_T_step_ahead} is obtained starting from the NARX equation \eqref{eq:narx}.

The NARX model structure requires the knowledge of the past inputs and the past measured outputs to obtain the predicted model output $y(t|\theta)$ (see \eqref{eq:narx_pred}). However, when simulating $T$ steps ahead, starting from time $t-1$, the measured outputs cannot be assumed to be known starting from time $t$ and onward. Hence, the classical approach is to replace them with past modeled outputs. This results in:
\begin{align}
    \hat{y}(t|t-1) &= f(u(t), \ldots, u(t-n_b), y(t-1),\ldots, y(t-n_a)), \label{eq:NARX_Appendix_1} \\
    \hat{y}(t+1|t-1) &= f(u(t+1), \ldots, u(t-n_b+1), \label{eq:NARX_Appendix_2}  \\
                     & \quad \quad \hat{y}(t|t-1), y(t-1), \ldots, y(t-n_a+1)), \nonumber \\
                     &= F_1(u(t+1), \ldots, u(t-n_b), \\ 
                     & \quad \quad  y(t-1), \ldots, y(t-n_a)) \nonumber
\end{align}
where $F_1(.)$ is obtained by substituting $\hat{y}(t|t-1)$ by \eqref{eq:NARX_Appendix_1} in \eqref{eq:NARX_Appendix_2}. Continued substitution for the prediction of the output samples further ahead in time results in \eqref{eq:NARX_T_step_ahead}.

\end{document}